\documentclass[11pt,letterpaper]{article}
\usepackage{geometry}
\usepackage{authblk}
\usepackage{times}
\usepackage{epsfig}
\usepackage{graphicx}
\usepackage{amsmath}
\usepackage{amssymb}
\usepackage{booktabs} 
\usepackage{comment}

\usepackage[breaklinks=true,bookmarks=false]{hyperref}

\begin{document}

\title{Machine Learning-based Automatic Graphene Detection with Color Correction for Optical Microscope Images}

\author[1]{Hui-Ying Siao}
\author[1]{Siyu Qi}
\author[1]{Zhi Ding}
\author[2]{Chia-Yu Lin}
\author[3]{Yu-Chiang Hsieh}
\author[3]{Tse-Ming Chen}
\affil[1]{Department of Electrical and Computer Engineering, University of California, Davis, CA 95616, USA}
\affil[2]{Department of Computer Science and Engineering, Yuan Ze University, Taoyuan City 32003, Taiwan}
\affil[3]{Department of Physics, National Cheng Kung University, Tainan 70101, Taiwan}


\date{}
\maketitle

\begin{abstract}
Graphene serves critical application and research purposes in various fields. 
However, fabricating high-quality and large quantities of graphene is time-consuming and it requires heavy human resource labor costs. 
In this paper, we propose a Machine Learning-based Automatic Graphene Detection Method with Color Correction (MLA-GDCC), a reliable and autonomous graphene detection from microscopic images. 
The MLA-GDCC includes a white balance (WB) to correct the color imbalance on the images, a modified U-Net and a support vector machine (SVM) to segment the graphene flakes. 
Considering the color shifts of the images caused by different cameras, we apply WB correction to correct the imbalance of the color pixels. 
A modified U-Net model, a convolutional neural network (CNN) architecture for fast and precise image segmentation, is introduced to segment the graphene flakes from the background. 
In order to improve the pixel-level accuracy, we implement a SVM after the modified U-Net model to separate the monolayer and bilayer graphene flakes. 
The MLA-GDCC achieves flake-level detection rates of 87.09\% for monolayer and 90.41\% for bilayer graphene, and the pixel-level accuracy of 99.27\% for monolayer and 98.92\% for bilayer graphene.
MLA-GDCC not only achieves high detection rates of the graphene flakes but also speeds up the latency for the graphene detection process from hours to seconds.
\end{abstract}

\section{Introduction}
Automatic graphene detection aims to collect the information for the material automatically in order to satisfy the increasing needs for the material from industry and academic research.
Graphene is a two-dimensional (2D) honeycomb lattice consisting of a single layer of carbon atoms. 
This 2D material has been found to have a broad range of applications in recent years, such as material science, physics, and device engineering owing to its unique physical characteristics  \cite{Novoselov2016,MaxC2007,Xia2009,Cao2018,Ohta2006,McCann_2013,Balandin2008,PhysRevLett.106.126802,Zhang2009, Lin2008}. Graphene can suspend millions of times its own weight, and is highly thermally conductive so there are limitless research and application purposes. Hence, researchers have been trying to fabricate high-quality and large quantities of graphene flakes in order to meet the high demands. Specifically, for research purposes, many research topics focus on the monolayer and bilayer graphene flakes \cite{Cao2018,Ramos2018}.

In order to obtain high quality 2D materials, research has shown that mechanical exfoliation is the state-of-the-art method \cite{huang2015exfoliation, Novoselov2004} to obtain the material for their research purposes. 
In general, there are three main steps to find and evaluate the thickness of graphene flakes in a research lab. First, graphene flakes are placed on a substrate with a mechanical exfoliation method. Second, the images of graphene flakes on top of the substrate are taken by an optical microscope. Last, after manually identifying the potential regions of interest, in a physics research lab, Raman spectroscopy\cite{Saito2011}, atomic force microscopy (AFM) \cite{Shearer_2016} and optical microscopy (OM) \cite{huang2015exfoliation} are the most common methods to manually identify the graphene layers. Due to its heavy dependence on human labor in the device fabrication process, high volume graphene production has been a challenge. 
In addition to the heavy cost of labor, the quality of the graphene flakes can deteriorate due to exposure to air and dirt. 
Therefore, in order to ensure the quality of the material, and to reduce the time and the labor costs, accelerating the fabrication process for the identification of the graphene flakes for making graphene-based devices has been a crucial research topic.

In recent years, many researchers applied artificial intelligence methods in order to identify graphene.
\cite{masubuchi2019classifying} applied clustering methods to identify the features and classify monolayer, bilayer and trilayer graphene. 
\cite{masubuchi2020deep} developed Mask-RCNN on the optical microscope 2D-material images to detect different materials, such as graphene and hBN, and classify the materials based on their thickness. 
However, a data-driven clustering analysis method and Mask-RCNN require a large dataset to train an accurate model. 
\cite{saito2019deep} implemented U-Net architecture to segment the monolayer and bilayer for both MoS$_{2}$ and graphene flakes. However, there is room for improvement on the false alarm rates (FAR) and the accuracy.
In common segmentation problems, the dataset provides multiple features, and the objects on the images usually have high contrast from the background. 
However, on the graphene microscopic images, there are limited features, and the pixel values of the background and the graphene flakes may overlap, resulting in high pixel-level FAR.
In addition, due to the color shift issues caused by different cameras, the modified U-Net fails to identify the regions of interest in our dataset without color correction. 

In order to develop a fast, reliable and autonomous graphene detection method, we propose a Machine Learning-based Automatic Graphene Detection Method with Color Correction (MLA-GDCC).
A MLA-GDCC contains a white balance (WB) method, a modified U-Net architecture, and a SVM algorithm to effectively segment the graphene flakes. 
To resolve the issue of the color shifts caused by different cameras, we apply a WB method to correct the color shifts on the input images. 
The WB method we implement in MLA-GDCC is the Gray world algorithm \cite{Buchsbaum1980, Chen2015,  Kim2002} which states that the average of all channels to be a gray image. 
The modified U-Net architecture can effectively segment the images of small datasets. Hence, in MLA-GDCC, the modified U-Net architecture is used to segment the regions of interest including monolayer and bilayer graphene from the background.
In order to improve the pixel-level accuracy, MLA-GDCC adds a SVM model as a pixel-level classifier to separate the monolayer and bilayer graphene. 

From the experiments, the pixel-level FAR is 0.51\% for monolayer graphene and 0.71\% for bilayer graphene. 
The pixel-level accuracy is 99.27\% for monolayer graphene and 98.92\% for bilayer graphene. 
By MLA-GDCC, the color shifts caused by the cameras is corrected.
The graphene flakes on each image can be identified in seconds to replace the labor-intensive process of finding the graphene flakes manually. 

\section{Related Works}
\begin{itemize}
\item{\bf U-Net for Image Segmentation:}
The U-Net architecture is one of the popular algorithms for semantic segmentation, otherwise known as pixel-based classification. It can efficiently segment the objects on the images with a very small training dataset. \cite{diagnostics10020110, Sevastopolsky2017, Yao2018, Vuola2019, YANG2020113419} There are various U-Net architectures modified in order to suit the dataset for better training processes. \cite{Khryashchev2019,Seo2020,Luo2018} 
The modified U-Net architecture in MLA-GDCC is designed to segment graphene, a nano-material, from the background.
In order for the modified U-Net to accurately detect and classify the objects on the images, multiple features are required. 
However, due to the features of the graphene dataset having only color pixel values, this is a limitation. Common features such as shapes and textures on the graphene microscopic images are irregular, and thus they cannot be used as features for the modified U-Net. The modified U-Net in our work can segment the regions of interest for our dataset and thus it efficiently blocks the background and the graphene flakes which are not part of the regions of interest. The regions of interest at the output of the modified U-Net contain the monolayer and the bilayer graphene. However, due to the overlapping of the pixel values between the monolayer and the bilayer graphene, simply applying a modified U-Net architecture cannot segment the monolayer and the bilayer graphene. Hence, a SVM is required in our system in order to differentiate the monolayer and bilayer graphene.

\item{\bf Pixel-Level Classifiers:}
A SVM is a supervised-learning algorithm using classification and regression to find the optimized data classification \cite{Bovolo2010,Sakthivel2014,Wang2011,Chapelle1999}. A SVM has been used as a pixel-level classifier for object segmentation in images. The authors of \cite{Wang2011} presented a color image segmentation method using a pixel-level SVM. Chapelle \textit{et al} \cite{Chapelle1999} showed that a SVM can generalize well on difficult image classification problems where the only features are high dimensional histograms. In 2014, Sakthivel \textit{et al} \cite{Sakthivel2014}, also proposed that by using a SVM trained by using Fuzzy C-Means (FCM) can effectively segment the color images. Unlike other works, in which datasets provide many trainable features\cite{Yao2009, HU2012133, Islam2017, Alquran2017, Hossain2018}. In our work, the features on the dataset are limited. The shapes and the textures for graphene are not specific and thus we cannot include these two features in our training process. The only feature that we can apply to our training process is the color pixel values of graphene.  In order to resolve this issue, we applied the SVM algorithm in MLA-GDCC as an image segmentation method to separate the monolayer and the bilayer graphene flakes from the regions of interest. Although the features in our dataset are very limited, the segmentation results for monolayer and bilayer graphene flakes show a high accuracy and a low FAR with the implementation of the SVM.
\end{itemize}



\section{Machine Learning-based Automatic Graphene Detection Method with Color Correction (MLA-GDCC)}
\begin{figure*}[h]
    \includegraphics[width=\linewidth]{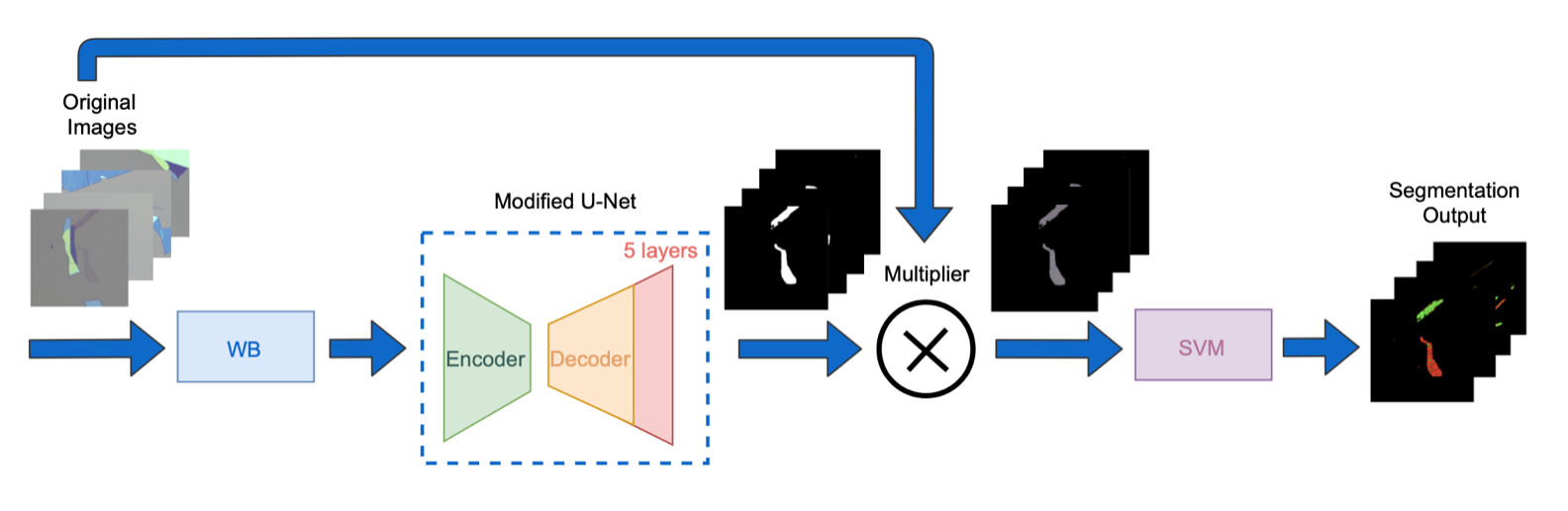}
    \caption{Complete process of the MLA-GDCC.}
    \label{fig:diagram}
\end{figure*}

Fig.~\ref{fig:diagram} shows the process of MLA-GDCC. 
First, we apply a WB method to the original images to correct the color shifts caused by different cameras. 
Second, the modified U-Net architecture is used to segment the regions of interest from the background in the white-balanced images. 
Third, a multiplier is implemented to mask the background from the white-balanced images. 
Last but not least, a SVM is implemented to classify the monolayer and bilayer graphene.

\subsection{White Balance Method (WB)}
Due to different types of the cameras used by different research groups to capture the graphene images, color shifts can happen on the images. 
Therefore, we implement a WB correction in the MLA-GDCC method to correct the color shifts. 
We implement a Gray world algorithm \cite{Buchsbaum1980, Chen2015,  Kim2002} as the WB method to make the average of all channels a gray image.
With the WB correction, models can accurately detect the graphene on the color-shifted microscopic images captured by different cameras.

\subsection{Modified U-Net architecture}
\begin{figure}[t] \label{compare_unet_m-unet.png}
    \includegraphics[width=\linewidth]{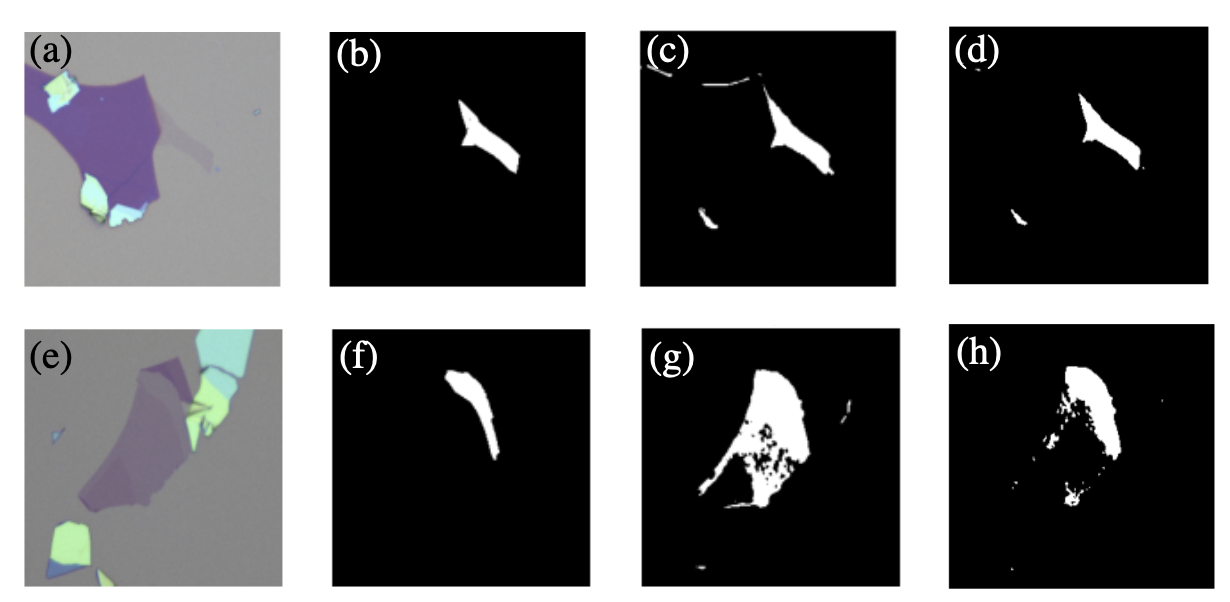}
    \caption{\textbf{(a)} and \textbf{(e)} are the original images. \textbf{(b)} and \textbf{(f)} are the GT masks. \textbf{(c)} and \textbf{(g)} are the outputs from the modified U-Net architecture. \textbf{(d)} and \textbf{(h)} are the outputs from the modified U-Net architecture with a threshold.}
    \label{fig:compare_unet&m-unet}
\end{figure}

In order to segment both the single layer and the bilayer graphene flakes from the background more accurately, we modify the traditional U-Net architecture \cite{Ronneberger2015} by adding 5 convolutional layers in the decoder to generate more training parameters from the images. 
The inputs of the modified U-Net are RGB microscopic images of graphene on the SiO$_{2}$ substrates, and all the input images are resized into 256 $\times$ 256 pixels.
The outputs of the modified U-Net are the detected masks containing both monolayer and bilayer graphene flakes, which provides pixel-level probability maps for the graphene devices. 
The output of the modified U-Net is shown in Fig. \ref{fig:compare_unet&m-unet}. (b) and (f) in Fig. \ref{fig:compare_unet&m-unet} show the ground truths (GT) of the graphene flakes for both monolayer and bilayer graphene. 


We use the weighted binary cross-entropy defined below as the loss for the modified U-Net, 
\begin{equation}\label{eq:loss}
\begin{split}
L =\frac{1}{N}\frac{1}{h\cdot w}\sum_{i,j}(k & \cdot y^{(g)}_{i,j}\log({\bar y}_{i,j})+ \\
&(1-y^{(g)}_{i,j})\log(1-{\bar y}_{i,j})),
\end{split}
\end{equation}

where $i$,$j$ are the pixel positions on each image, $k$ represents the positive weight. Due to the data imbalance of the number of pixels values between the background and the graphene flakes, positive weight is added here to increase the weight of the positive samples, which are graphene pixel values. ${\bar y}_{i,j}$ is the output of the modified U-Net, $N$ is the number of the images in the dataset, $h$ is the height, $w$ is the width and $y^{(g)}_{i,j}$ is the GT. 

\begin{equation}
{\bar y}_{i,j}=\theta(w_{i,j,m,z}\cdot x_{z,m}+b_{i,j}),
\label{eq:activation}
\end{equation}

In Eq. \ref{eq:activation}, $\theta$ stands for the activation function, here we implement a sigmoid function as the activation function to confine the output values among [0,1]. $w_{i,j,m,z}$ is the weight, $m$ and $z$ are the input dimension indexes, $x_{z,m}$ stands for the input of the neural network, $b_{i,j}$ is the bias.   


\subsection{SVM: Color Pixels Classifier}
\label{sect:method2}
\begin{figure}[t] \label{modified-unet.png}
    \includegraphics[width=\linewidth]{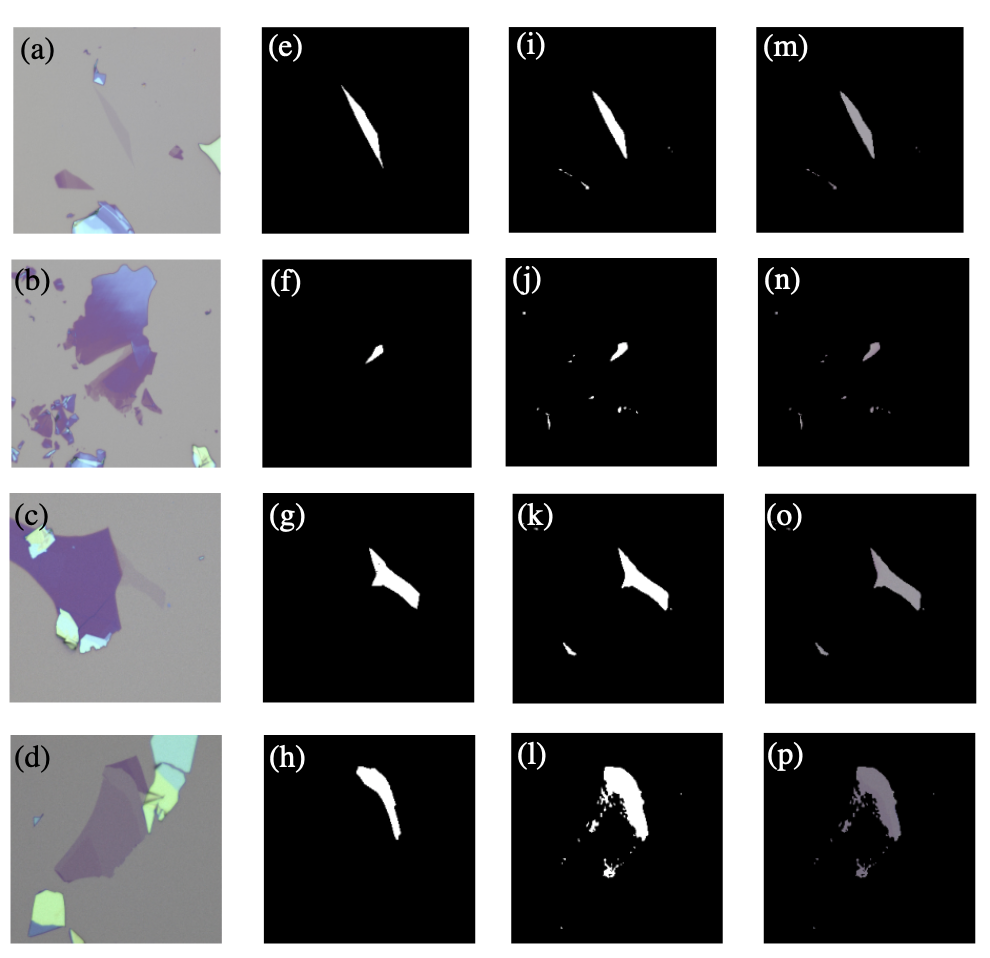}
    \caption{\textbf{(a)-(d)} are the original images. \textbf{(e)-(h)} are the GT masks. \textbf{(i)-(l)} are the outputs from the modified U-Net. \textbf{(m)-(p)} are the outputs from the multiplier.}
    \label{fig:modified-unet}
\end{figure}

A multiplier is implemented to mask the background in the white-balanced images as shown in Fig. \ref{fig:modified-unet}. 
Only the regions of interest on the white-balanced images are left at the output of the multiplier. 
The limited features on the graphene dataset prevent the modified U-Net from classifying the monolayer and the bilayer graphene. 
Therefore, we use a SVM to identify and classify the monolayer and bilayer graphene flakes more accurately. 
SVM is a machine learning algorithm used to find the hyper-plane based on the dataset characteristics, this algorithm aims to maximize the margin of the hyper-plane and the data points in a linear or a nonlinear case.

A SVM has two criterias including the empirical error minimization and the control of model complexity. 
With the two criteria, the cost function of the model can be minimized. The empirical error minimization reduces the errors during the training to optimize the training result.
The control of the model complexity can prevent overfitting by controlling the flexibility of the function. 
The pixel intensity indicates the number of layers of the graphene flakes.

Here we extract the RGB pixel values separately and assign the pixel values to the corresponding label $y_{i,j}$ indicating the layers of the graphene flakes. The features for the SVM algorithm include the mean values RGB pixel values and the mean RGB pixel values of the mode background pixel values. 
By using the detection masks from the U-Net, we use the histogram to select the peak value of the background for each R, G and B channel. Next, we average the three background RGB pixel values to obtain the mean RGB pixel value as one of the two features for the SVM algorithm. Since different images are composed of different mean pixel values, choosing the background pixel value as one of the features is necessary to allow the SVM to identify the graphene flakes more accurately.
Given by the input dataset $x$, the discriminant function for a linear SVM can be defined by
\begin{equation}
\begin{split}
{f(x)} =\sum^{N}_{i=1}w\langle x,x_{i}\rangle +b,  
\label{eq:linear_SVM}
\end{split}
\end{equation}

where $\omega$ is a vector normal to the hyper-plane and $b$ is a bias constant. However, the linear SVM cannot find the optimized hyper-plane for the graphene flake classification. According to the Cover's theorem \cite{Cover1965}, we can adopt a nonlinear SVM to map $x$ onto the higher dimensional space with the nonlinear mapping function, $\phi (x)$. Therefore, the discriminant function for nonlinear SVM can be written as  
\begin{equation}
\begin{split}
{f(x)} =\sum^{N}_{i=1}y_{i}\langle\phi(x),\phi(x_{i})\rangle +b,  
\label{eq:nonlinear_SVM}
\end{split}
\end{equation}
where $y_{i}$ is the label of the training pattern $x_{i}$, and $N$ is the total number of training samples. 
The discriminate function in Eq.\ref{eq:nonlinear_SVM} is derived by minimizing the cost function defined as follows
\begin{equation}
\begin{split}
{C(\omega,\beta_{i})} =\frac{1}{2}||w^{2}||+p\sum^{N}_{i=1}\beta_{i},
\label{eq:cost}
\end{split}
\end{equation}
where $\beta_{i}$ denotes the distance to the correct margin $\beta_{i}\geq$ 0.
with the constraints are as follows:
\begin{equation}\label{eq:constraint1}
\begin{split}
{\beta_{i}} \geq 1-y_{i}[\omega\cdot\phi(x_{i})+b], \: i=1,2,...,N
\end{split}
\end{equation}
where 
\begin{equation}\label{eq:constraint2}
\begin{split}
{\beta_{i}}\geq 0 \:\rm and\: C>0
\end{split}
\end{equation}

where $p$ is the regularization parameter, and $\beta_{i}$ are non-negative slack variables which are used to resolve the noisy and nonlinearly separable data.

And thus the decision function becomes
\begin{equation}\label{eq:decision}
\begin{split}
{g(x)} =sgn[f(x)]
\end{split}
\end{equation}

According to the Mercer’s theorem \cite{Cristianini2000, Mercer1909} the inner product in Eq.~\ref{eq:linear_SVM} can be replaced by a Kernel function that is chosen for the dataset distribution.
\begin{equation}\label{eq:replaceInner}
\begin{split}
{<\phi(x),\phi(x_{i})>} = K(x_{i},x_{j})
\end{split}
\end{equation}

With the replacement of the Kernel function as shown in Eq.~\ref{eq:replaceInner}, we can rewrite the discriminate function as the following equation:
\begin{equation}\label{eq:new_discriminat}
\begin{split}
{f(x)} =\sum^{N}_{i=1}\alpha_{i}\omega_{i}K(x,x_{i})+b,
\end{split}
\end{equation}
where $K(x,x_{i})$ represents the Kernel function. In our case, we select Gaussian Kernel to fit our dataset. The Gaussian Kernel can be defined as follows:
\begin{equation}\label{eq:Gaussian}
\begin{split}
{K(x_{i},x_{j})} = exp(-||x_{i}-x_{j}||^2/2\sigma^2)
\end{split}
\end{equation}

With the implementation of the SVM algorithm, we can discriminate two classes in our dataset. The two classes here are the monolayer and bilayer graphene mean RGB pixel values.  The SVM algorithm does not require a lot of training data. Although the training dataset is small, which in our case is 246 images, the model achieves high detection rates of the monolayer and bilayer graphene flakes.

\section{Experiments}

\label{sect:results}
\subsection{Datasets and Implementation Environments} 
The dataset in this work is the graphene microscopic images obtained using mechanical exfoliation on top of the SiO$_{2}$/Si wafers. The dataset includes 246 images for training and 57 images for testing. The height and width of the images are both 256 pixels. The GTs are labeled manually by the authors using labelbox.com. The GPU which is used to train the models in our work is a single 12GB NVIDIA Tesla K80 GPU.

\subsection{Evaluation Metrics} 
We apply the evaluation metrics of Eqs.~\ref{eq:precision}-~\ref{eq:FAR} to evaluate the results from the SVM. In the equations, P stands for graphene pixels and N stands for background pixels for the pixel-level evaluation metrics. For the flake-level DR, TP stands for graphene flakes correctly detected, and FN stands for graphene flakes failed to be detected by the model.

\begin{equation}
\begin{split}
\rm{Precision(\%)} =\frac{TP}{TP+FP}.  
\label{eq:precision}
\end{split}
\end{equation}

\begin{equation}
\begin{split}
\rm{F1\; score(\%)} =\frac{2\cdot Precision \cdot Recall}{Precision+Recall}.  
\label{eq:f1-score}
\end{split}
\end{equation}

\begin{equation}
\begin{split}
\rm{Recall(DR)(\%)} =\frac{TP}{TP+FN}. 
\label{eq:recall}
\end{split}
\end{equation}

\begin{equation}
\begin{split}
\rm{Accuracy(\%)} =\frac{TP+TN}{TP+FN+TN+FP}.  
\label{eq:accuracy}
\end{split}
\end{equation}

\begin{equation}
\begin{split}
\rm{False\; alarm\; rate(\%)} =\frac{FP}{FP+TN}.  
\label{eq:FAR}
\end{split}
\end{equation}

\subsection{The Accuracy of MLA-GDCC}
The training process of the modified U-Net is 450 epochs, with a learning rate of 0.001, and the positive weight $k$ of 200. 
We adopt the Adam optimizer \cite{Kingma2014} for training the modified U-Net. 
The modified U-Net is used to separate the graphene flakes from the background. 
The pixel-level detection rate of the modified U-Net on the test dataset is 97\%, and the false alarm rate of the modified U-Net is 3\%. 

\begin{figure}[t] \label{ROC.png}
    \includegraphics[width=\linewidth]{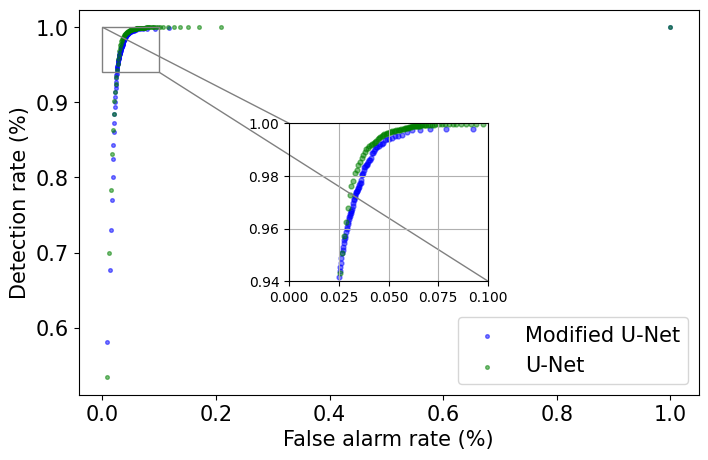}
    \caption{ROC comparison between the U-Net and the modified U-Net architecture.}
    \label{fig:ROC2}
\end{figure}

We also compare the receiver operating characteristic (ROC) curves of the U-Net with and without the additional 5 layers in Fig. \ref{fig:ROC2}.
At the knee point of the curves we observe an improvement of the implementation of the additional 5 convolutional layers added to the decoder. 
The detection rate at the knee point is increased compared to the U-Net architecture. With a higher detection rate at the output using the modified U-Net, we achieved an improvement of up to 0.7\% in a detection rate of the modified U-Net compared to the original U-Net.

\begin{figure}[t]
    \includegraphics[width=\linewidth]{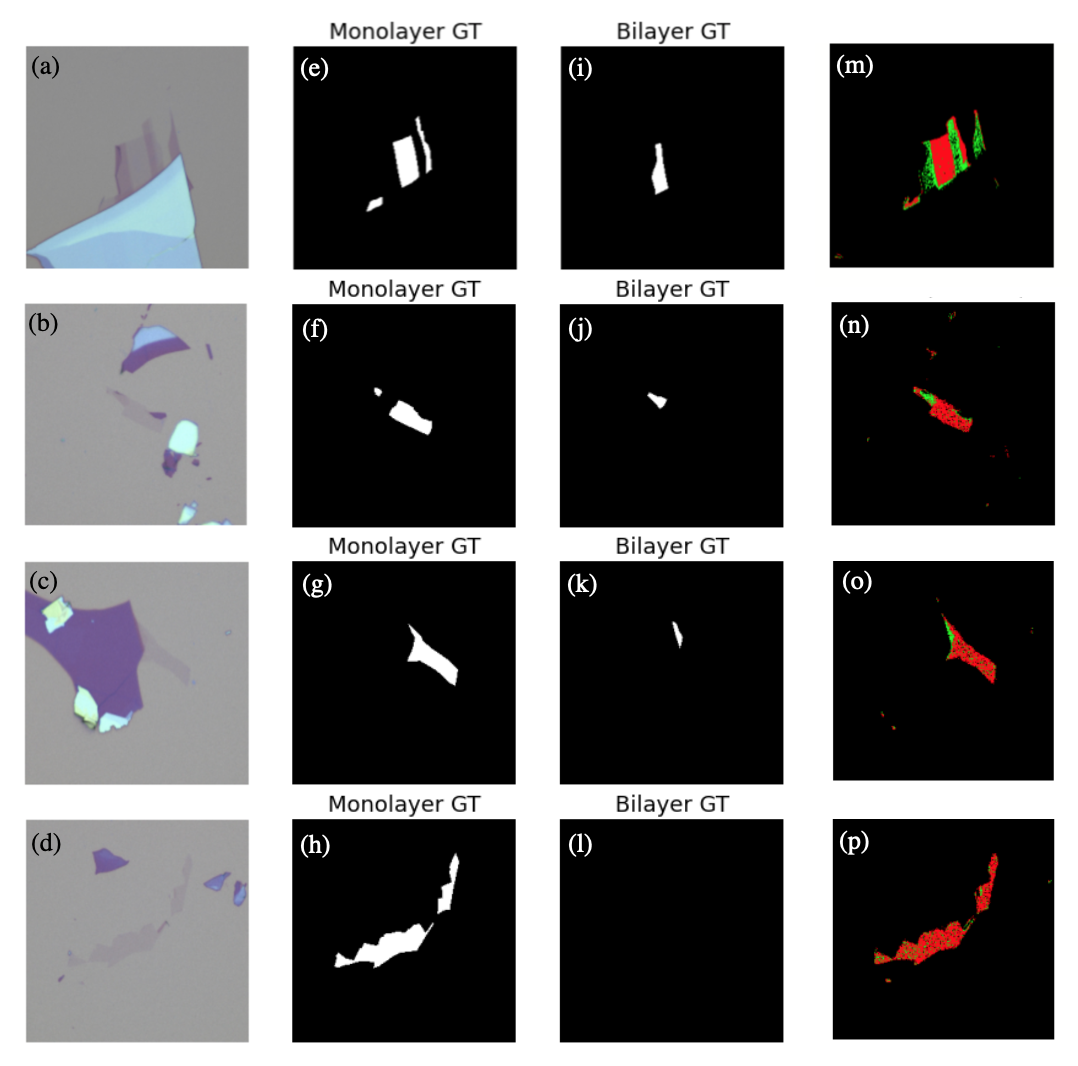}
    \caption{\textbf{(a)-(d)} are the original images. \textbf{(e)-(h)} are the ground truth images of the monolayer graphene. \textbf{(i)-(l)} are the ground truth images of the bilayer graphene.   \textbf{(m)-(p)} are the detected monolayer(green area) and bilayer(red area) graphene.}
    \label{fig:svmm}
\end{figure}

A multiplier is used to multiply the detected masks and the original images. 
Therefore, at the output of the multiplier, the detected masks cover the background and only the graphene regions of interest are left. 
The output from the multiplier are the input for the SVM. With the implementation of the SVM algorithm, the monolayer and the bilayer graphene can be separated at the output of the SVM, as shown in Fig. \ref{fig:svmm}. 
The flake-level detection rates are 87.09\% for monolayer and 90.41\% for bilayer graphene.

The pixel-level evaluation metrics for MLA-GDCC calculated using from Eq. \ref {eq:precision} to Eq. \ref{eq:accuracy} can be found in Table \ref {table:dr}. 
The precision is 51.01\% for monolayer and 70.37\% for bilayer graphene.
The F1 score is 59.03\% for monolayer and 75.38\% for bilayer graphene.
The recall or detection rate is 70.05\% for monolayer and 81.16\% for bilayer graphene.
The accuracy is 99.27\% for monolayer and 98.92\% for bilayer graphene.


\subsection{Ablation Study of White Balance Method} 
\begin{figure}[t]
    \includegraphics[width=\linewidth]{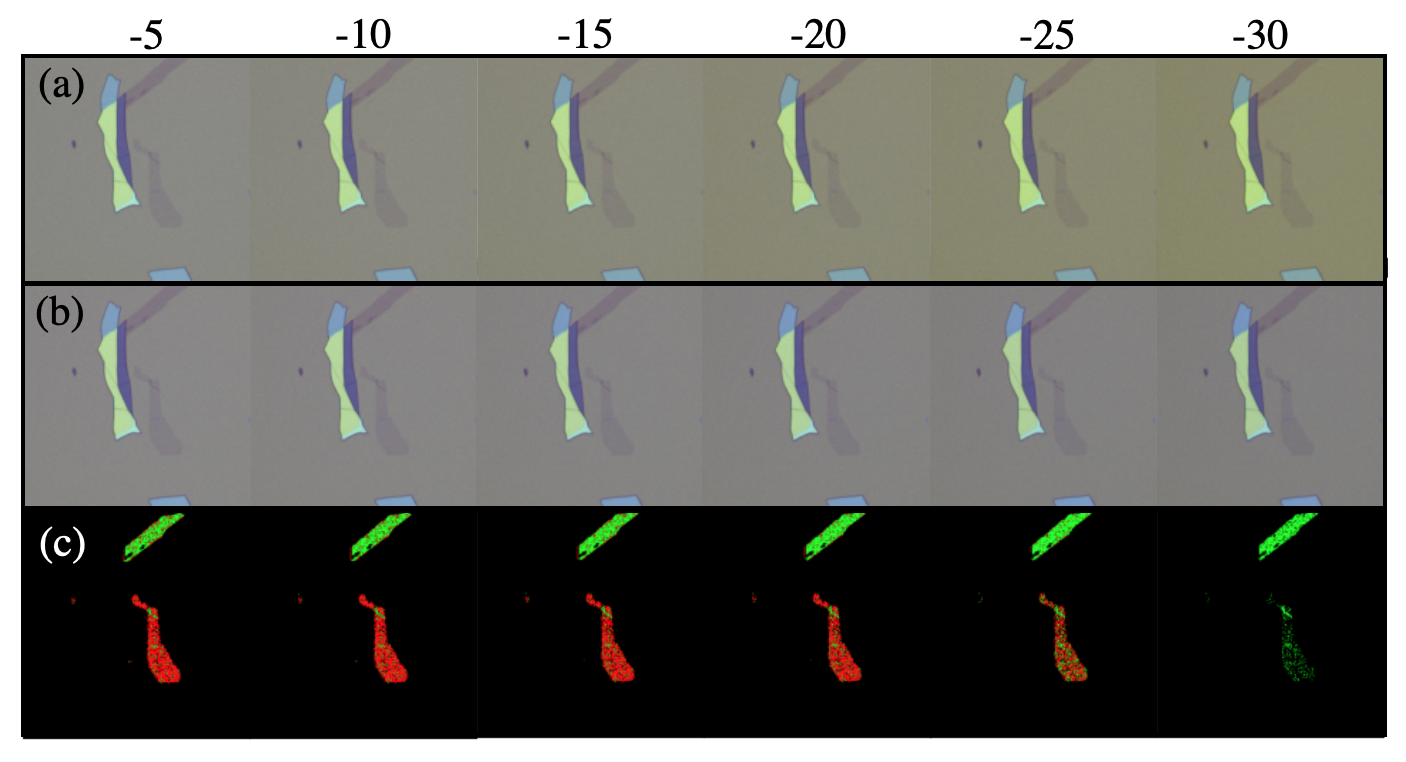}
    \caption{\textbf{(a)} shows the original images with the pixel values in the blue channel subtracted with the corresponding values on the images.  \textbf{(b)} shows the images after applying white balance on (a). \textbf{(c)} shows the detection results from the SVM.}
    \label{fig:wb_minus}
\end{figure}

\begin{figure}[h]
    \includegraphics[width=\linewidth]{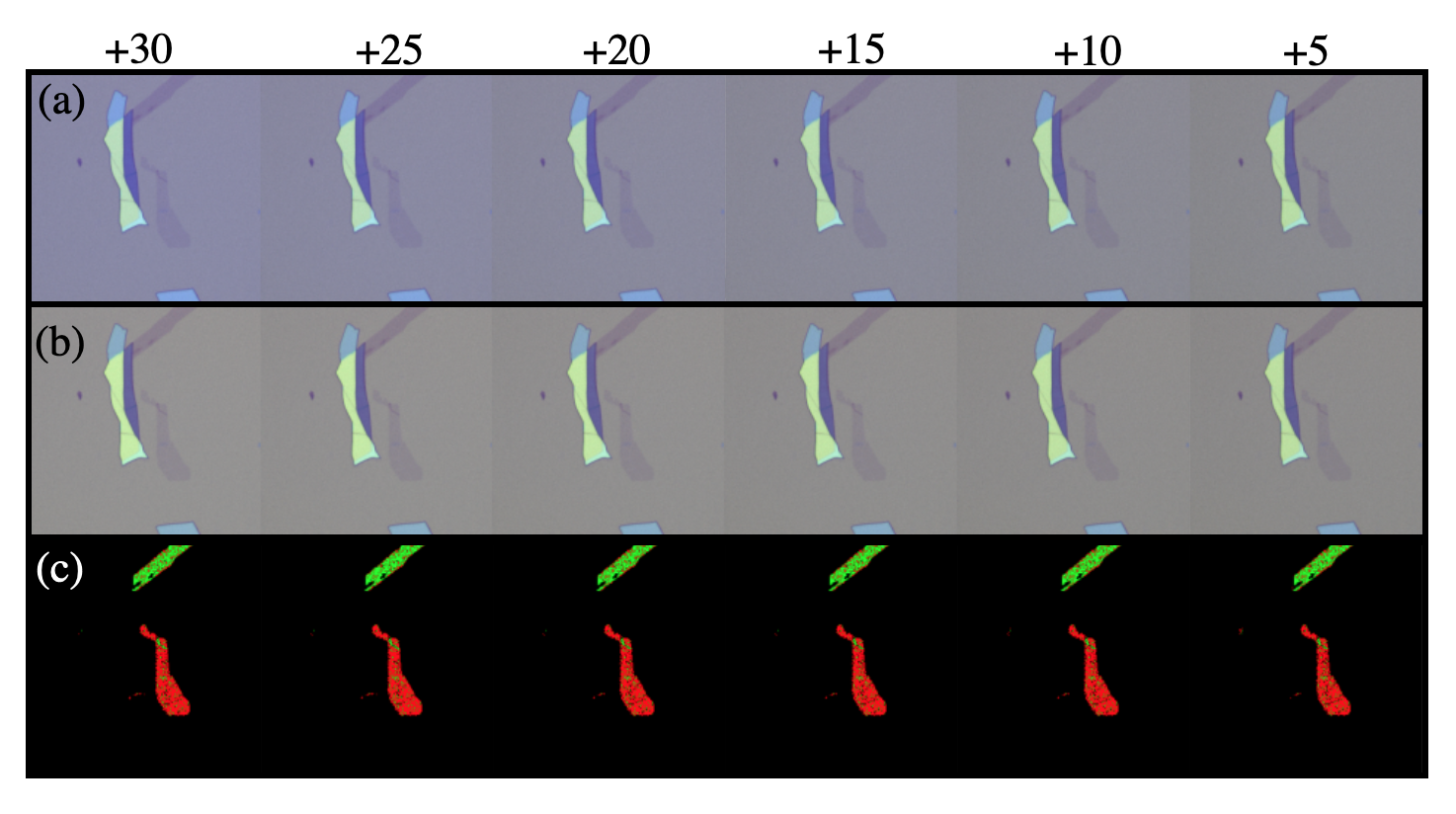}
    \caption{\textbf{(a)} shows the original images with pixel values in the blue channel added with the corresponding values on the images. \textbf{(b)} shows the images after applying white balance on (a). \textbf{(c)} shows the detection results from the SVM.
    \label{fig:wb_plus}}
\end{figure}

\begin{figure}[t] 
    \includegraphics[width=\linewidth]{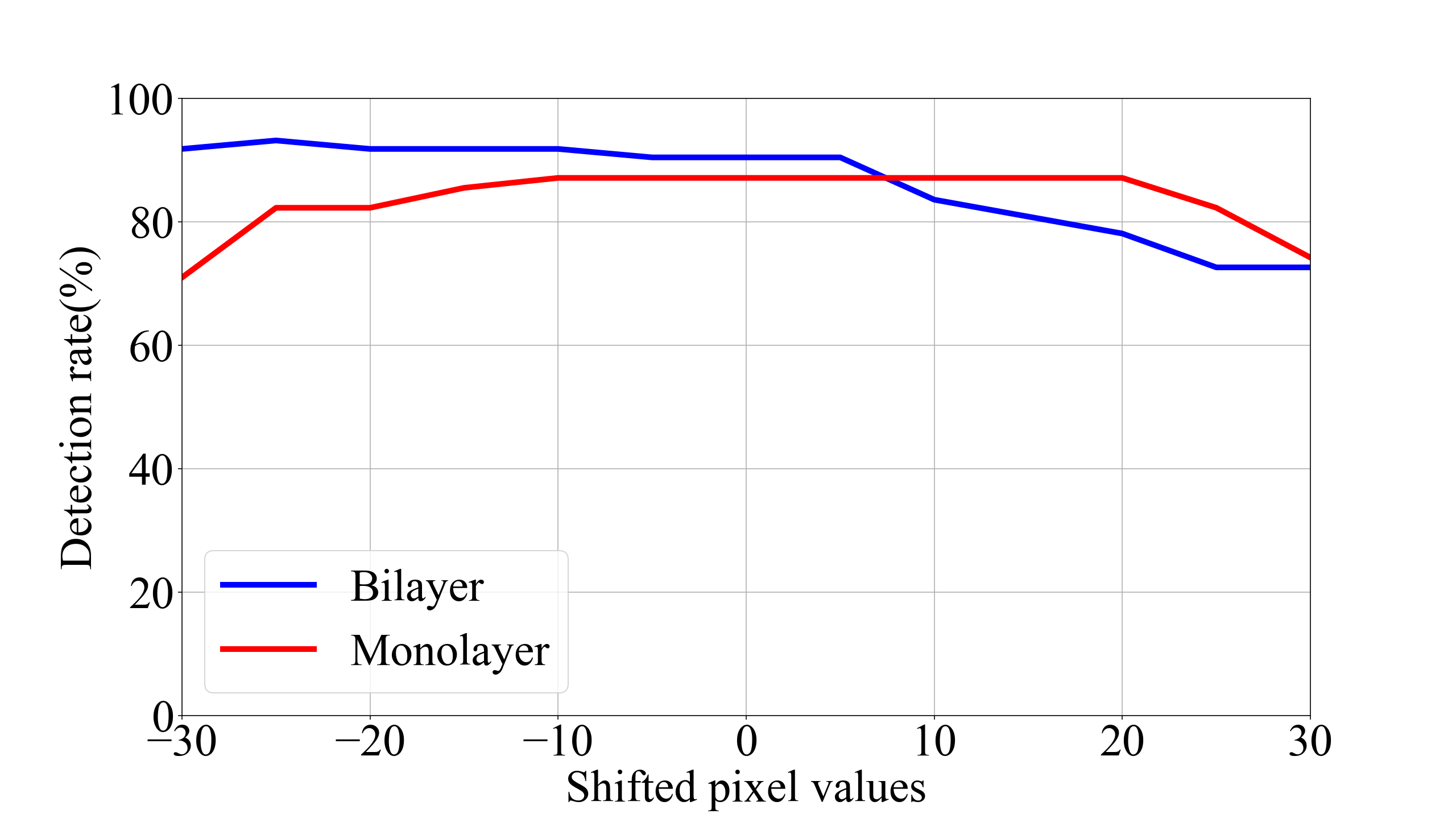}
    \caption{Detection rates of the input images shifted with different pixel values.}
    \label{fig:detection_plot}
\end{figure}

With the implementation of the Gray world algorithm, the results show that the SVM model can detect the graphene flakes and identify the layers with high detection rates. The detection results of the color-shifted images are shown in Fig.\ref{fig:wb_minus} and Fig.\ref{fig:wb_plus}. 
In both Fig.\ref{fig:wb_minus} and Fig.\ref{fig:wb_plus}, part \textbf{(a)} shows the images shifted with different numbers of pixel values in the blue channel. Part \textbf{(b)} shows the application of the Gray world algorithm on the images in part \textbf{(a)}. Part \textbf{(c)} shows the output from the MLA-GDCC. The numerical results from the output of the MLA-GDCC can be found in Table \ref{table:dr-flake}. We use the numerical results in Table \ref{table:dr-flake} to plot Fig.\ref{fig:detection_plot} and demonstrate the detection rates with different shifted pixel values of the original images. With the WB correction applied to the color-shifted images, the MLA-GDCC remains at high detection rates. The result shows how the robustness and the reliability of the color correction method applied in the MLA-GDCC can deal with the potential color-shift caused by different cameras.

\section{Conclusions}
\label{sect:conclusions}
In this paper, a MLA-GDCC method is introduced to automatically, precisely and rapidly classify monolayer and bilayer graphene flakes. 
With the implementation of the Gray world algorithm, the color shifts on the microscopic images caused by different cameras are resolved. We implement a modified U-Net architecture and a SVM algorithm to achieve high detection rates of graphene flakes on the microscopic images. The MLA-GDCC allows us to segment the graphene layers with high flake-level detection rates of 87.09\% for monolayer and 90.41\% for bilayer graphene, and high pixel-level accuracy of 99.27\% for monolayer and 98.92\% for bilayer graphene.

{\small
\bibliographystyle{ieee_fullname}
\bibliography{egbib}

\begin{thebibliography}{10}\itemsep=-1pt

\bibitem{Alquran2017}
H. Alquran, I.~A. Qasmieh, A.~M. Alqudah, S. Alhammouri, E. Alawneh, A.
  Abughazaleh, and F. Hasayen.
\newblock The melanoma skin cancer detection and classification using support
  vector machine.
\newblock pages 1--5, 2017.

\bibitem{Balandin2008}
A. Balandin, S. Ghosh, W. Bao, I. Calizo, D. Teweldebrhan, F. Miao, and C.~N.
  Lau.
\newblock Superior thermal conductivity of single-layer graphene.
\newblock {\em Nano Letters}, 8(3):902--907, 2008.

\bibitem{Bovolo2010}
F. Bovolon, L. Bruzzone, and L. Carlin.
\newblock A novel technique for subpixel image classification based on support
  vector machine.
\newblock {\em IEEE TRANSACTIONS ON IMAGE PROCESSING}, 19:2983--2999, 2010.

\bibitem{Buchsbaum1980}
G. Buchsbaum.
\newblock A spatial processor model for object colour perception.
\newblock 310:1--26, 1980.

\bibitem{Cao2018}
Y. Cao, V. Fatemi1, S. Fang, K. Watanabe, T. Taniguchi, E. Kaxiras, and P.
  Jarillo-Herrero1.
\newblock Unconventional superconductivity in magic-angle graphene
  superlattices.
\newblock 556:43--50, 2018.

\bibitem{Chapelle1999}
O. Chapelle, P. Haffner, and V.~N. Vapnik.
\newblock Support vector machines for histogram-based image classification.
\newblock {\em IEEE TRANSACTIONS ON NEURAL NETWORKS}, 10:1055--1064, 1999.

\bibitem{Chen2015}
G. Chen and X. Zhang.
\newblock A method to improve robustness of the gray world algorithm.
\newblock pages 250--255, 2015.

\bibitem{Cover1965}
T.~M. Cover.
\newblock Geometrical and statistical properties of systems of linear
  inequalities with application in pattern recognition.
\newblock 14:326–--334, 1965.

\bibitem{Cristianini2000}
N. Cristianini and J. Shawe-Tayloi.
\newblock An introduction to support vector machines.
\newblock 2000.

\bibitem{diagnostics10020110}
P. Gadosey, Y. Li, E. Agyekum, T. Zhang, Z. Liu, P. Yamak, and F. Essaf.
\newblock Sd-unet: Stripping down u-net for segmentation of biomedical images
  on platforms with low computational budgets.
\newblock {\em Diagnostics}, 10(2), 2020.

\bibitem{Hossain2018}
S. Hossain, R. Mou, M. Hasan, and and M.~Razzak S.~Chakraborty.
\newblock Recognition and detection of tea leaf's diseases using support vector
  machine.
\newblock {\em 2018 IEEE 14th International Colloquium on Signal Processing and
  Its Applications (CSPA)}, pages 150--154, 2018.

\bibitem{HU2012133}
J. Hu, D.Li, Q. Duan, Y. Han, G. Chen, and X. Si.
\newblock Fish species classification by color, texture and multi-class support
  vector machine using computer vision.
\newblock {\em Computers and Electronics in Agriculture}, 88:133--140, 2012.

\bibitem{huang2015exfoliation}
Y. Huang, E. Sutter, N.~N. Shi, J.~Zheng~T. Yang, D. Englund, H-J Gao, and P.
  Sutter.
\newblock Reliable exfoliation of large-area high-quality flakes of graphene
  and other two-dimensional materials.
\newblock {\em ACS Nano}, 9(11):10612–--10620, 2015.

\bibitem{Islam2017}
M. Islam, Anh Dinh, K. Wahid, and P. Bhowmik.
\newblock Detection of potato diseases using image segmentation and multiclass
  support vector machine.
\newblock pages 1--4, 2017.

\bibitem{Mercer1909}
JMercer.
\newblock Functions of positive and negative type and their connection with the
  theory of integral equations.
\newblock 209:415–--446, 1909.

\bibitem{Khryashchev2019}
V. Khryashchev, R. Larionov, A. Ostrovskaya, and A. Semenov.
\newblock Modification of u-net neural network in the task of multichannel
  satellite images segmentation.
\newblock {\em 2019 IEEE East-West Design Test Symposium (EWDTS)}, pages 1--4,
  2019.

\bibitem{Kim2002}
Y. Kim, J-S Lee, A. Morales, and S-J Ko.
\newblock A video camera system with enhanced zoom tracking and auto white
  balance.
\newblock {\em IEEE Transactions on Consumer Electronics}, 48:428--434, 2002.

\bibitem{Kingma2014}
D.~P. Kingma and J. Ba.
\newblock Adam: A method for stochastic optimization.
\newblock 2014.

\bibitem{MaxC2007}
M.~C. Lemme, T.~J. Echtermeyer, M. Baus, and H. Kurz.
\newblock A graphene field-effect device.
\newblock {\em IEEE Electron Device Letters}, 28:282--284, 2007.

\bibitem{Lin2008}
Y-M Lin and P. Avouris.
\newblock Strong suppression of electrical noise in bilayer graphene
  nanodevices.
\newblock {\em Nano Letters}, 8(8):2119--2125, 2008.

\bibitem{PhysRevLett.106.126802}
A. Luican, G. Li, A. Reina, J.~Kong andR. R.~Nair, K.~S. Novoselov, A.~K. Geim,
  and E.~Y. Andrei.
\newblock Single-layer behavior and its breakdown in twisted graphene layers.
\newblock {\em Phys. Rev. Lett.}, 106:126802, Mar 2011.

\bibitem{Luo2018}
L. Luo, D. Chen, and D. Xue.
\newblock Retinal blood vessels semantic segmentation method based on modified
  u-net.
\newblock pages 1892--1895, 2018.

\bibitem{masubuchi2019classifying}
S. Masubuchi and T. Machida.
\newblock Classifying optical microscope images of exfoliated graphene flakes
  by data-driven machine learning.
\newblock {\em npj 2D Materials and Applications}, 3(1):1--7, 2019.

\bibitem{McCann_2013}
Edward McCann and Mikito Koshino.
\newblock The electronic properties of bilayer graphene.
\newblock {\em Reports on Progress in Physics}, 76(5):056503, 2013.

\bibitem{Novoselov2004}
K.~S. Novoselov, A.~K. Geim, S.~V. Morozov, D. Jiang, Y. Zhang, S.~V. Dubonos,
  I.~V. Grigorieva, and A.~A. Firsov.
\newblock Electric field effect in atomically thin carbon films.
\newblock {\em Science}, 306(5696):666--669, 2004.

\bibitem{Novoselov2016}
K.~S. Novoselov, A. Mishchenko, A. Carvalho, and A.~H.~Castro Neto.
\newblock 2d materials and van der waals heterostructures.
\newblock {\em Science}, 353(6298):aac9439, 2016.

\bibitem{Ohta2006}
T. Ohta, A. Bostwick, T. Seyller, Karsten Horn, and Eli Rotenberg.
\newblock Controlling the electronic structure of bilayer graphene.
\newblock {\em Science}, 313:951--954, 2006.

\bibitem{Ramos2018}
J.~G. G.~S. Ramos, T.~C. Vasconcelos, and A.~L.~R. Barbosa.
\newblock Spin-to-charge conversion in 2d electron gas and single-layer
  graphene devices.
\newblock 123, 2018.

\bibitem{Ronneberger2015}
O. Ronneberger, P.Fischer, and T. Brox.
\newblock U-net: Convolutional networks for biomedical image segmentation.
\newblock 9351:234--241, 2015.

\bibitem{masubuchi2020deep}
Eisuke S.~Masubuchi, and~Watanabe, Yuta Seo, Shota Okazaki, Takao Sasagawa,
  Kenji Watanabe, Takashi Taniguchi, and Tomoki Machida.
\newblock Deep-learning-based image segmentation integrated with optical
  microscopy for automatically searching for two-dimensional materials.
\newblock {\em npj 2D Materials and Applications}, 4(1):1--9, 2020.

\bibitem{Saito2011}
R. Saito, M. Hofmann, G. Dresselhaus, A. Jorio, and M.~S. Dresselhaus.
\newblock Raman spectroscopy of graphene and carbon nanotubes.
\newblock {\em Advances in Physics}, 60(3):413--550, 2011.

\bibitem{saito2019deep}
Y. Saito, K. Shin, K. Terayama, S. Desai, M. Onga, Y. Nakagawa, Y. Itahashi, Y.
  Iwasa, M. Yamada, and K. Tsuda.
\newblock Deep-learning-based quality filtering of mechanically exfoliated 2d
  crystals.
\newblock {\em npj Computational Materials}, 5(1):1--6, 2019.

\bibitem{Sakthivel2014}
K. Sakthivel, R. Nallusamy, and C. Kavitha.
\newblock Color image segmentation using svm pixel classification image.
\newblock {\em IEEE TRANSACTIONS ON IMAGE PROCESSING}, 8:1924--1930, 2014.

\bibitem{Seo2020}
H. Seo, C. Huang, M. Bassenne, R. Xiao, and L. Xing.
\newblock Modified u-net (mu-net) with incorporation of object-dependent high
  level features for improved liver and liver-tumor segmentation in ct images.
\newblock {\em IEEE Transactions on Medical Imaging}, 39(5):1316--1325, 2020.

\bibitem{Sevastopolsky2017}
A. Sevastopolsky.
\newblock Optic disc and cup segmentation methods for glaucoma detection with
  modification of u-net convolutional neural network.
\newblock 27:618–624, 2017.

\bibitem{Shearer_2016}
C. Shearer, A. Slattery, A. Stapleton, J. Shapter, and C. Gibson.
\newblock Accurate thickness measurement of graphene.
\newblock {\em Nanotechnology}, 27(12):125704, feb 2016.

\bibitem{Vuola2019}
A.~O. Vuola, S.~U. Akram, and J. Kannala.
\newblock Mask-rcnn and u-net ensembled for nuclei segmentation.
\newblock pages 208--212, 2019.

\bibitem{Wang2011}
X-Y. Wang, T. Wang, and J. Bu.
\newblock Color image segmentation using pixel wise support vector machine
  classification.
\newblock {\em Pattern Recognition}, 44:777--787, 2011.

\bibitem{Xia2009}
F. Xia, T. Mueller, Y-M Lin, A. Valdes-Garcia, and P. Avouris.
\newblock Ultrafast graphene photodetector.
\newblock {\em Nature Nanotechnology}, 4:839–--843, 2009.

\bibitem{YANG2020113419}
Y. Yang, C. Feng, and R. Wang.
\newblock Automatic segmentation model combining u-net and level set method for
  medical images.
\newblock {\em Expert Systems with Applications}, 153:113419, 2020.

\bibitem{Yao2009}
Q. Yao, Z. Guan, Y. Zhou, J. Tang, Y. Hu, and B. Yang.
\newblock Application of support vector machine for detecting rice diseases
  using shape and color texture features.
\newblock {\em 2009 International Conference on Engineering Computation}, pages
  79--83, 2009.

\bibitem{Yao2018}
W. Yao, Z. Zeng, C.Lian, and H. Tang.
\newblock Pixel-wise regression using u-net and its application on
  pansharpening.
\newblock 312:364–371, 2018.

\bibitem{Zhang2009}
Y. Zhang, and C.~Girit T-T~Tang, Z. Hao, M. Martin, A. Zettl, M. Crommie, Y.
  Shen, and F. Wang.
\newblock Direct observation of a widely tunable bandgap in bilayer graphene.
\newblock {\em Nature}, 459:820–823, 2009.

\end{thebibliography}
}

\begin{table*}[t]
	\caption{Pixel-level evaluation metrics} 
	\label{table:dr}
	\centering
	\begin{tabular}[]{c|c|c|c|c}
		\toprule
        & precision(\%) & F1 score(\%) & recall(DR)(\%) & accuracy(\%)\\
        \hline
        Monolayer & 51.01&59.03&70.05&99.27\\
        \hline
        Bilayer & 70.37 &75.38&81.16&98.92\\
		\bottomrule
	\end{tabular}

    \vspace{4pt}
	\caption{Flake-level detection rates changing with the shifted pixel values}
	\label{table:dr-flake}
	\centering
	\begin{tabular}[]{c|c|c|c|c|c|c|c|c|c|c|c}
		\toprule
        Shifted pixel values &-25&-20&-15&-10&-5&0&5&10&15&20&25\\
        \hline
        Monolayer DR(\%)& 82.25&82.25&85.48&87.09&87.09&87.09&87.09&87.09&87.09&87.09&82.25\\
        \hline
        Bilayer DR(\%)& 93.15&91.78&91.78&91.78&90.41&90.41&90.41&83.56&80.82&78.08&72.6\\
		\bottomrule
	\end{tabular}

    \vspace{4pt}
	\caption{Pixel-level detection rates and false alarm rates changing with the shifted pixel values}
	\label{table:dr-2}
	\centering
	\begin{tabular}[]{c|c|c|c|c|c|c|c|c|c|c|c}
		\toprule
        Shifted pixel values &-25&-20&-15&-10&-5&0&5&10&15&20&25\\
        \hline
        Monolayer DR(\%)& 
        62.45&63.7&65.31&67.08&67.75&70.05&71.33&72.49&71.4&68.07&59.89\\
        \hline
        Bilayer DR(\%)& 
        81.63&81.11&80.92&81.4&80.81&81.16&80.15&74.32&74.54&68.55&66.46\\
        \hline
        Monolayer FAR(\%)& 
        0.41&0.45&0.47&0.48&0.51&0.51&0.56&0.68&0.7&0.77&0.8\\
        \hline
        Bilayer FAR(\%)& 
        0.75&0.73&0.66&0.72&0.71&0.71&0.69&0.68&0.74&0.76&0.76\\
		\bottomrule
	\end{tabular}
\end{table*}

\end{document}